\documentclass[aps,pre,twocolumn,groupedaddress]{revtex4-1}
\usepackage{graphicx}
\usepackage{mathrsfs}
\usepackage{amsmath}
\begin{document}

\preprint{}

\title{Singular-value decomposition using quantum annealing}

%%%%%%%%%%%%%%%%%%%%%%%%%Author Names%%%%%%%%%%%%%%%%%%%%%%%

\author{Yoichiro Hashizume}
 \email{hashizume@rs.tus.ac.jp}
\author{Takashi Koizumi}
\author{Kento Akitaya}
\author{Takashi Nakajima}
 \email{nakajima@rs.tus.ac.jp}
\affiliation{Department of Applied Physics, Tokyo University of Science, Tokyo 125-8585, Japan}

\author{Masuo Suzuki}
 \email{masuo.suzuki@riken.jp}
\affiliation{Computational Astrophysics Laboratory, RIKEN, 2-1 Hirosawa, Wako, Saitama, 351-0198, Japan}

\author{Soichiro Okamura}
 \email{sokamura@rs.kagu.tus.ac.jp}
\affiliation{Department of Applied Physics, Tokyo University of Science, Tokyo 125-8585, Japan}

\begin{abstract}
In the present study, we demonstrate how to perform, using quantum annealing, the singular value decomposition and the principal component analysis.
Quantum annealing gives a way to find a ground state of a system, while the singular value decomposition requires the maximum eigenstate.
The key idea is to transform the sign of the final Hamiltonian, and the maximum eigenstate is obtained by quantum annealing.
Furthermore, the adiabatic time scale is obtained by the approximation focusing on the maximum eigenvalue.
\end{abstract}


%%%%%%%%%%%%%%%%%%%%%%%%%Keywords for PhysicaA%%%%%%%%%%%%%%%%%%%%%%%
%\begin{keyword}
%singular value decomposition
%\sep
%principal component analysis
%\sep
%quantum annealing
%\sep
%image analysis
%\end{keyword}
%%%%%%%%%%%%%%%%%%%%%%%%%%%%%%%%%%%%%%%%%%%%%%%%%%%%%%%%%%%%%%%%%%

\maketitle

\section{\label{intro} Introduction}
Quantum annealing~[1-10] is a useful way to estimate the ground state of a system.
Here and after we denote the target Hamiltonian as $\mathcal{H}_1$.
In the famous quantum annealing schemes~[1-10], the time dependent Hamiltonian $\mathcal{H}(t)$ is introduced as
\begin{equation}
\mathcal{H}(t)=\left( 1-\frac{t}{T} \right)\mathcal{H}_0+\frac{t}{T}\mathcal{H}_1 \label{eq1-1}
\end{equation}
using the initial system $\mathcal{H}_0$.
Here the characteristic time scale $T$ is assumed to be long enough for the Schr\"{o}dinger equation
\begin{equation}
i\hbar \frac{\partial }{\partial t}|\psi(t)\rangle = \mathcal{H}(t)|\psi(t)\rangle \label{eq1-2}
\end{equation}
to be regarded to describe the adiabatic process, namely
\begin{equation}
\mathcal{H}(t)|\psi(t)\rangle = E(t)|\psi(t)\rangle, \label{eq1-3}
\end{equation}
where the eigenvalue $E(t)$ corresponds to the energy of the total system (\ref{eq1-1}).
Starting from such an initial condition as the ground state $|\psi(0)\rangle=|\phi_0\rangle$ of the initial system $\mathcal{H}_0$, we can obtain the ground state of the system $\mathcal{H}_1$ at $t=T$.
This is a simple outline of quantum annealing.

Quantum annealing is often applied to estimate a ground state~[10-16] in such complex systems as spin-glasses, and these studies are applied to non-deterministic polynomial (NP) problems such as the traveling salesman problem~[11-14].
On the other hand, recently, information technology becomes more important and attention is paid to the statistical physics of information~[17-19].
Especially, the scheme to treat the big data from aspects of statistical physics will develop in the near future.
Actually, for example, Kurihara {\it et al.} studied network clustering using quantum annealing~\cite{15,16}.
Thus, it may be important to analyze directly the big data from those kinds of viewpoints.

In the case of data analysis, the principal component analysis is often used to find the trends shown in big data.
Furthermore, the principal component analysis is equivalent to the singular value decomposition.
Once we can perform the singular value decomposition using quantum annealing and clarify its mechanism, quantum annealing provides an efficient way of data analysis. 

Additionally, from the viewpoint of applications for fundamental problems of quantum physics, the singular value decomposition gives a useful way to study quantum states.
For example, to find the entanglement between subsystems $\mathcal{H}_{\text{A}}$ and $\mathcal{H}_{\text{B}}$ of the total Hamiltonian $\mathcal{H}_{\text{tot}}=\mathcal{H}_{\text{A}}+\mathcal{H}_{\text{A}}$, we have to obtain a reduced density matrix.
To obtain the reduced density matrix, Schmidt decomposition and singular value decomposition are used.
In other cases, to perform the density matrix renormalization group (DMRG) method, the singular value decomposition is necessary.
Thus, it is also useful to show a possible way to perform the singular value decomposition. 

In the present study, we show how to perform the singular value decomposition using quantum annealing and we investigate the mechanism.
In the next section, we make a brief introduction of the singular value decomposition from the viewpoint of an application of quantum annealing.
In Sec. \ref{sec3}, we introduce the key idea to perform the singular value decomposition by quantum annealing, and demonstrate it explicitly by some information matrices.
We analyze the mechanism of the present way by using approximated eigenvalue distributions and a series expansion, in Secs. \ref{sec4} and \ref{sec5}, respectively.
Summary and discussions are included in Sec. \ref{sec6}.

\section{\label{sec2} Singular value decomposition}
In this section, we make a brief introduction of the singular value decomposition.
We consider an $m\times n$ matrix $A$ for natural numbers $m$ and $n$.
The matrix $G$ is defined as
\begin{equation}
G=A^{\dagger}A, \label{eq2-1}
\end{equation}
which is an $n\times n$ Hermitian matrix.
The eigenvectors $\{|v_j\rangle\}$ ($j=0,1,2,\dots$) and the eigenvalues $\{\lambda_j\}$ of the matrix $G$ satisfy the equation
\begin{equation}
G|v_j\rangle=\lambda _j |v_j\rangle. \label{eq2-2}
\end{equation}
Here the eigenvalues $\{\lambda_j\}$ are real because $G$ is Hermitian.
In addition, using the inequality
\begin{equation}
0\leq ||A|v_j\rangle ||^2 =\langle v_j|A^{\dagger}A|v_j\rangle =\lambda_j, \label{eq2-2.5} 
\end{equation}
it is easily found that the eigenvalue $\lambda_j$ is positive.

If we define the vector $|u_j\rangle$ as
\begin{equation}
|u_j\rangle =\frac{1}{\sqrt{\lambda_j}}A|v_j\rangle, \label{eq2-3}
\end{equation}
the vector $AA^{\dagger}|u_j\rangle$ is derived as
\begin{equation}
AA^{\dagger}|u_j\rangle = \frac{1}{\sqrt{\lambda_j}}AG|v_j\rangle = \sqrt{\lambda_j}A|v_j\rangle = \lambda_j |u_j\rangle. \label{eq2-4}
\end{equation}
Thus, the vector $|u_j\rangle$ is an eigenvector of the matrix $AA^{\dagger}$ with the eigenvalue $\lambda_j$ including $m$ components.

Using the vectors $\{|u_j \rangle \}$, $\{|v_j \rangle \}$ and the eigenvalues $\{\lambda_j\}$, the matrix $A$ is decomposed as
\begin{equation}
A=\sum_{j}A|v_j\rangle \langle v_j| = \sum_{j}\sqrt{\lambda_j}|u_j\rangle \langle v_j|. \label{eq2-5}
\end{equation}
This decomposition is so called ``singular value decomposition'', and the weights $\{\sqrt{\lambda_j}\}$ are called ``singular values''.
To use the singular value decomposition for data analysis, the component with larger singular values are more important.
Thus, we put the eigenvalues $\{\lambda_j\}$ in order as $\lambda_0\geq \lambda_1\geq \lambda_2\geq \dots$.
In this case, $|v_0\rangle$ is called ``first principal component'', $|v_1\rangle$ is called ``second principal component'', and so on.
This is nothing but the principal component analysis~\cite{20}.
In many cases, even though it is not necessary, the data matrix $A=(a_{ij})$ is often normalized so as to satisfy the relations
\begin{equation}
\bar{a}_j=\frac{1}{m}\sum_{k=1}^{m}a_{kj}=0, \label{eq2-6}
\end{equation} 
and
\begin{equation}
\sigma_j = \frac{1}{m}\sum_{k=1}^{m}\left( a_{kj}-\bar{a}_j \right)^2=1. \label{eq2-7}
\end{equation}
When the data matrix $A$ is normalized, the matrix $G\equiv A^{\dagger}A$ corresponds to the variance-covariance matrix.
Then, the principal component analysis corresponds to the eigenvalue analysis of the variance-covariance matrices.

As shown in Eq.(\ref{eq2-5}), the singular value decomposition and the principal component analysis require the eigenstate with larger eigenvalues, because the larger singular values strongly contribute to the original data matrix $A$ with the weights $\sqrt{\lambda_j}$.
Thus, for the data analysis, it is enough to find the first- and second-principal components.
In the present study, we try to find the first-principal component $|v_0\rangle$ (whose eigenvalue is the largest one), using quantum annealing.

\section{\label{sec3} Demonstration of singular value decomposition by quantum annealing}
To perform the singular value decomposition, we need to obtain the largest eigenvalue of the matrix $G=A^{\dagger}A$ for the information matrix $A$.
However, the quantum annealing method yields the ground state of the target Hamiltonian $\mathcal{H}_1$.
Then, we put the target Hamiltonian $\mathcal{H}_1$ as $-G$, namely
\begin{equation}
\mathcal{H}_1=-G=-A^{\dagger}A. \label{eq3-1}
\end{equation}
Thus, the total Hamiltonian $\mathcal{H}(t)$ is defined as 
\begin{equation}
\mathcal{H}(t)=-\frac{t}{T}G+\left(1-\frac{t}{T}\right)\mathcal{H}_0, \label{eq3-2}
\end{equation}
using Eq.(\ref{eq1-1}).
Generally, the matrix $G$ often includes the non-zero off-diagonal elements which play the role of quantum effects of the Schr\"{o}dinger equation
\begin{align}
i\hbar \frac{\partial }{\partial t}|\psi(t)\rangle &=\mathcal{H}(t)|\psi(t)\rangle \notag
\\
&=\left[ -\frac{t}{T}G+\left(1-\frac{t}{T}\right)\mathcal{H}_0 \right] |\psi(t)\rangle. \label{eq3-3}
\end{align}
The initial Hamiltonian $\mathcal{H}_0$ to treat the present problem is defined as the diagonal matrix
\begin{align}
\mathcal{H}_0&=-\left( \Lambda_0+\Lambda \right)|\phi_0\rangle \langle \phi_0|+\Lambda \notag
\\
&=\begin{pmatrix}
-\Lambda_0 & 0& 0 &\cdots
\\
0& \Lambda & 0 &\cdots
\\
0&0& \Lambda &\cdots
\\
\cdots
\end{pmatrix} \label{eq3-4}
\end{align}
with some positive parameters $\Lambda_0$ and $\Lambda$.
Using the bases $\langle \phi_0| = (1,0,0,\dots), \langle \phi_1|=(0,1,0,\dots)$ and so on, the initial Hamiltonian $\mathcal{H}_0$ yields the ground state $|\phi_0\rangle $ with the eigenvalue $-\Lambda_0$, and the degenerated exited states $|\phi_1\rangle, |\phi_2\rangle, \dots $ with the same eigenvalue $\Lambda$.
The initial state $|\psi(0)\rangle $ is assumed as the ground state $|\phi_0\rangle $, namely
\begin{equation}
|\psi(0)\rangle =|\phi_0\rangle. \label{eq3-5}
\end{equation}
The time development of the state vector $|\psi(t)\rangle$ is obtained by
\begin{equation}
|\psi(t+dt)\rangle =|\psi(t)\rangle +\frac{1}{i\hbar}\mathcal{H}(t)|\psi(t)\rangle dt \label{eq3-6}
\end{equation}
from the Schr\"{o}dinger equation (\ref{eq3-3}).
Finally, if quantum annealing works well, we can expect that the first-principal component $|v_0 \rangle$ is obtained as $|\psi(T)\rangle = |v_0 \rangle$ at the time $t=T$.

In this section, on the basis of the above assumption, we demonstrate two examples, namely, a small data matrix and an image data matrix.
Especially, recent studies clarified the importance of the image analysis using the singular value decomposition from the viewpoint of critical phenomena.
The present demonstrations may be interesting from such a point of view.

\subsection{Singular value decomposition of two-dimensional data}
In the present section, we show a simple example to demonstrate the singular value decomposition of a small matrix using quantum annealing.
We consider a league table showing the records of two persons (students) and their three exams as a normalized form.
Such an information matrix is denoted as
\begin{equation}
A=\begin{pmatrix}
a_{11}&a_{12}
\\
a_{21}&a_{22}
\\
a_{31}&a_{32}
\end{pmatrix}
=\begin{pmatrix}
-0.69&-0.68
\\
-0.023&0.73
\\
0.72&-0.043
\end{pmatrix}, \label{eq3-7}
\end{equation}
where the parameters $\{ a_{ij} \}$ are normalized to satisfy the relations (\ref{eq2-6}) and (\ref{eq2-7}).
The variance-covariance matrix $G$ is obtained as
\begin{equation}
G=A^{\dagger}A=\begin{pmatrix}
1&0.43
\\
0.43&1
\end{pmatrix}.\label{eq3-8}
\end{equation}
In the present case, the eigenvalues $\lambda_0$ and $\lambda_1$ ($\lambda_0>\lambda_1$), and the eigenvectors $|v_0\rangle$ and $|v_1\rangle$ are easily obtained as
\begin{equation}
\lambda_0 =1.43 {\text{ and }} \lambda_1 =0.57, \label{eq3-9} 
\end{equation}
and
\begin{equation}
|v_0\rangle =\frac{1}{\sqrt{2}}\begin{pmatrix} 1\\1 \end{pmatrix}  {\text{ and }} |v_1\rangle =\frac{1}{\sqrt{2}}\begin{pmatrix} 1\\-1 \end{pmatrix}, \label{eq3-10}
\end{equation}
respectively.
We try to obtain the above first-principal component $|v_0\rangle$ using quantum annealing. 

To perform the quantum annealing, the initial Hamiltonian $\mathcal{H}_0$ and its ground state $|\phi_0\rangle$ are defined as
\begin{equation}
\mathcal{H}_0=\begin{pmatrix}-1&0\\0&1\end{pmatrix} {\text{ and }} |\phi_0 \rangle =\begin{pmatrix} 1\\0 \end{pmatrix} \label{eq3-11}
\end{equation}
by Eq. (\ref{eq3-4}), where we assume $\Lambda_0=\Lambda=1$.
Thus, the time dependent Hamiltonian $\mathcal{H}(t)$ is expressed as
\begin{equation}
\mathcal{H}(t)=-\frac{t}{T}\begin{pmatrix}1&0.43\\0.43&1\end{pmatrix}\epsilon +\left(1-\frac{t}{T}\right) \begin{pmatrix}-1&0\\0&1\end{pmatrix}\epsilon, \label{eq3-12}
\end{equation}
where the parameter $\epsilon$ denotes an ``energy constant'' introduced to clarify the unit of the energy.
Assuming the large time scale $T$, the Shr\"{o}dinger equation can be approximated as
\begin{equation}
i\hbar \frac{\partial }{\partial t}|\psi(t)\rangle =\mathcal{H}(t)|\psi(t)\rangle \simeq E(t)|\psi(t)\rangle, \label{eq3-13}
\end{equation}
as discussed in Sec. \ref{intro}.
The ground state energy $E(t)$ is obtained as
\begin{align}
E(t)&=\frac{\epsilon}{T}\left( -t-0.5\sqrt{4.74t^2-8Tt+4T^2} \right) \notag
\\
&\xrightarrow[t\to T]{ }-1.43\epsilon, \label{eq3-14}
\end{align}
and the state vector $|\psi(T)\rangle$ is also obtained from Eq.(\ref{eq3-6}) as
\begin{equation}
|\psi(T) \rangle =\begin{pmatrix} 0.707 \\ 0.707 \end{pmatrix} \simeq \frac{1}{\sqrt{2}} \begin{pmatrix} 1 \\ 1 \end{pmatrix}, \label{eq3-15}
\end{equation}
where we have assumed $T=10^3\hbar/\epsilon$.
The state $|\psi(T)\rangle$ is nothing but the first-principal component $|v_0\rangle$ which is the eigenstate of $G$ with the {\it largest} eigenvalue $\lambda_0=1.43$.
As shown in the present demonstration, we can perform the singular value decomposition by quantum annealing using the Hamiltonian (\ref{eq3-2}).
To obtain the second largest eigenstate, namely the second-principal component, we rewrite Eq. (\ref{eq2-5}) as
\begin{align}
A&=\sqrt{\lambda_0}|u_0\rangle\langle v_0|+\sqrt{\lambda_1}|u_1\rangle\langle v_1|+\cdots \notag
\\
\Leftrightarrow A-\sqrt{\lambda_0}|u_0\rangle\langle v_0|&=\sqrt{\lambda_1}|u_1\rangle\langle v_1|+\cdots.\label{eq3-16}
\end{align}
As shown in Eq. (\ref{eq3-16}), under the redefinition of the information matrix $A$ as $A\to A-\sqrt{\lambda_0}|u_0\rangle\langle v_0|$, we can find the second-principal component $|v_1\rangle$ as the first-principal component.
Then, we obtain the arbitrary spectrum of the singular value decomposition in a step-by-step manner.

As shown in this demonstration, our algorithm offers no guarantee of speeding up over a classical algorithm even for determining one principal component.
Determining all others one by one is as classical as it could be.

\subsection{Image analysis by quantum annealing}
\begin{figure}[hbpt]
\includegraphics[width=2.5in]{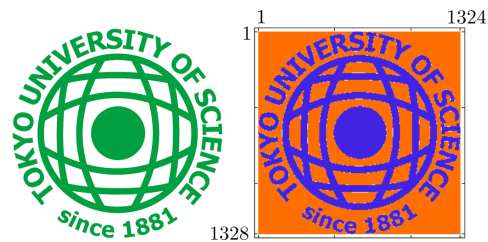}
\caption{\label{fig1} (Color online) The logo of Tokyo University of Science (left-hand side). The binary image of the logo of TUS (right-hand side) is regarded as the information matrix $A$.}
\end{figure}

\begin{figure}[hbpt]
\includegraphics[width=3in]{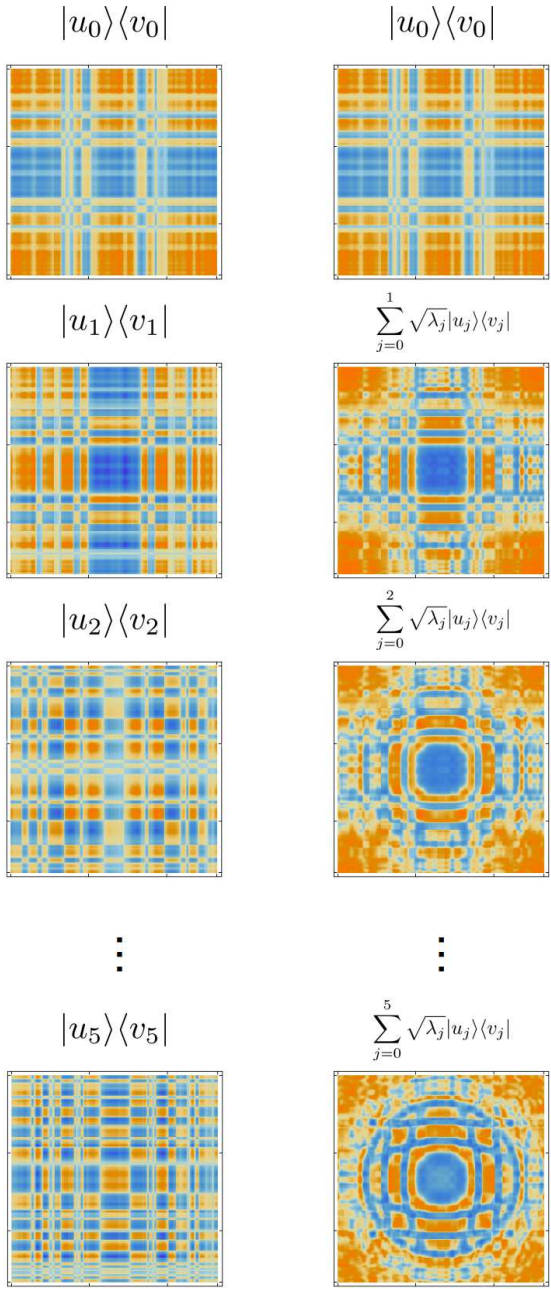}
\caption{\label{fig2} (Color online) Singular value decomposition of the logo of TUS given by the quantum annealing method. All figures show decomposed $1328 \times 1324$ matrices. The left-hand sides show each component $|u_j\rangle \langle v_j|$ while the right-hand sides show the summation $\sum \sqrt{\lambda_j}|u_j\rangle \langle v_j|$.}
\end{figure}

Recently, Matsueda {\it et al}. proposed a new idea of the way to image analysis based on the singular value decomposition~[21-25].
These previous studies suggested that the singular values $\sqrt{\lambda_j}$ decay in a power law or exponentially reflecting the correlation scales shown in the image~[21-23].
In the present section, we try to perform the singular value decomposition of an image by quantum annealing.

The original test image is the logo of Tokyo University of Science (TUS) as shown in Fig.\ref{fig1}.
Clearly, we can find highly symmetric and several scales are included in this figure.
Here we use the binary $(1,-1)$ data of the logo, and obtain the $1328\times 1324$ matrix $A$.  
Using the present scheme as discussed in the previous sections, we obtain the decomposed image as shown in Fig. \ref{fig2}.
The left-hand sides of Fig. \ref{fig2} show each component $|u_j\rangle \langle v_j|$ while the right-hand sides show the summation $\sum \sqrt{\lambda_j}|u_j\rangle \langle v_j|$.
As shown in Fig. \ref{fig2}, the singular value decomposition splits the data matrix into layers characterized by correlation scales as was pointed out in the previous studies~[24,25] by Matsueda {\it et al}.
In the present case, we also find that quantum annealing works well enough to perform the singular value decomposition of the image data.

\section{\label{sec4} Analytical explanation of the present method}
In the present section, we study the mechanism of the present method to perform the singular value decomposition by quantum annealing.
The time-dependent Hamiltonian $\mathcal{H}(t)$ is denoted as
\begin{equation}
\mathcal{H}(t)=\mathcal{H}(x)=-xG+(1-x)\mathcal{H}_0, \label{eq4-1}
\end{equation}
where the parameter $x$ denotes $x=t/T$.
Assuming the adiabatic process with large $T$, the Schr\"{o}dinger equation (\ref{eq3-3}) yields the equation
\begin{equation}
\mathcal{H}(x)|\psi(x)\rangle=E(x)|\psi(x)\rangle \label{eq4-2}
\end{equation}
using the eigenvalue $E(x)$.
The initial condition of $|\psi(x)\rangle$ is assumed as $|\psi(0)\rangle =|\phi_0\rangle$ which is the ground state of the initial Hamiltonian (\ref{eq3-4}).
We assume the simple case $\Lambda_0=\Lambda$ in the initial Hamiltonian (\ref{eq3-4}).
The target matrix $G$ satisfies the equation
\begin{equation}
G|v_j\rangle =\lambda_j |v_j\rangle, \label{eq4-3}
\end{equation}
for all principal components $\{|v_j \rangle \}$, where the eigenvalues $\{\lambda_j\}$ are put in order as
\begin{equation}
\lambda_0 \geq \lambda_1 \geq \cdots \geq 0. \label{eq4-4}
\end{equation}

As shown in the previous studies~[21-25], the largest eigenvalue $\lambda_0$ is much larger than the second largest one $\lambda_1$.
Thus, we assume here the approximation $\lambda_1/\lambda_0\simeq 0$.
In this prediction, the state vector $|\psi(x)\rangle $ is approximately expressed as
\begin{equation}
|\psi(x)\rangle \simeq a(x)|v_0\rangle +b(x)|\phi_0\rangle \label{eq4-5}
\end{equation}
using the functions $a(x)$ and $b(x)$ as the coefficients.
Then, Eq. (\ref{eq4-2}) yields
\begin{widetext}
\begin{align}
& \left[ -xG+(1-x)\mathcal{H}_0 \right] \left[a(x)|v_0\rangle +b(x)|\phi_0\rangle \right] =E(x) \left[ a(x)|v_0\rangle +b(x)|\phi_0\rangle \right] \notag
\\
\Leftrightarrow &
\begin{cases}
\left[-\lambda_0 x+(1-x)\Lambda_0\right] a(x)-\lambda_0 x\langle v_0|\phi_0 \rangle b(x)=E(x)a(x)
\\
-2\Lambda_0 (1-x)\langle v_0|\phi_0 \rangle a(x) +\left[-(1-x)\Lambda_0\right]  b(x)=E(x)b(x).
\end{cases}
\label{eq4-6}
\end{align}
Therefore, the energy $E(x)$ is obtained as
\begin{equation}
E(x)=\frac{\Lambda_0}{2}\left[ -x K\pm \sqrt{x^2K^2+4(1-x)^2+4x(1-x)(2\alpha ^2-1)K} \right], \label{eq4-7}
\end{equation}
\end{widetext}
where the parameters $K$ and $\alpha$ are defined as $K=\lambda_0/\Lambda_0$ and $\alpha=\langle v_0|\phi_0 \rangle=\langle \phi_0|v_0 \rangle$, respectively.
As shown in Eq. (\ref{eq4-7}), the initial eigenvalues $\pm \Lambda_0$ are expressed as $E(0)=\pm \Lambda_0$, while the final eigenvalues $E(1)$ become $-\lambda_0$ and $0$.
One of the final eigenvalues, $E(1)=0$, should be $E(1)=-\lambda_1$ rigorously.
This is due to the approximation $\lambda_1/\lambda_0\simeq 0$.

\begin{figure}[hbpt]
\includegraphics[width=3in]{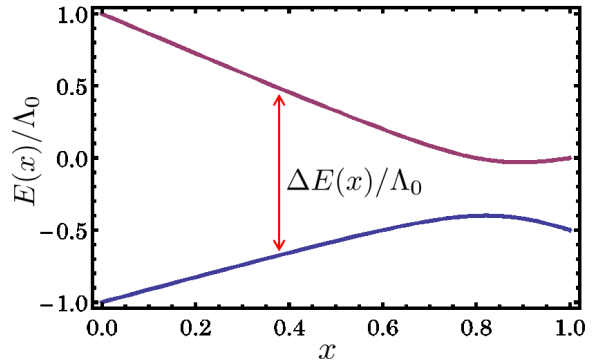}
\caption{\label{fig3} (Color online) Eigenvalue properties of the ground state and the excited state. The parameters $K$ and $\alpha$ are assumed as $K=\alpha=0.5$. $\Delta E(x)$ denotes the energy gap.}
\end{figure}

\begin{figure}[hbpt]
\includegraphics[width=3in]{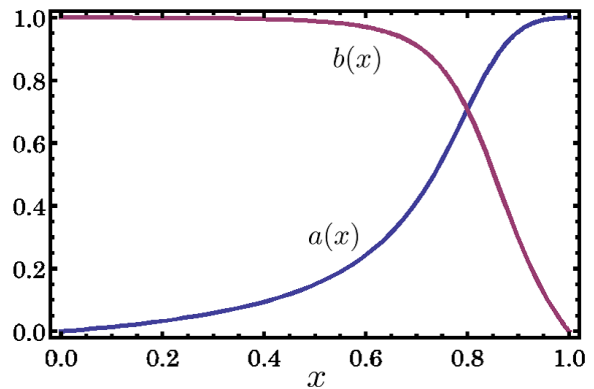}
\caption{\label{fig4} (Color online) The functions $a(x)$ and $b(x)$.}
\end{figure}

We show the $x$ dependence of $E(x)$, $a(x)$, and $b(x)$ in Figs. \ref{fig3} and \ref{fig4}.
Both of the numerical calculations shown in these figures, the parameters $K$ and $\alpha$ are assumed as $K=\alpha=0.5$.
Figure \ref{fig3} shows the energy profiles through the quantum annealing process.
The energy gap $\Delta E(x)$ has the minimum value
\begin{equation}
{\text{min}}\left[\Delta E(x) \right] =4\Lambda_0\sqrt{\frac{K^2\alpha^2 (1-\alpha^2 )}{(2+K)^2-8K\alpha^2}}. \label{eq4-8}
\end{equation}
As is well known~\cite{26}, the appropriate time scale $T$ satisfies the relation
\begin{equation}
T\sim \frac{\Lambda_0 \hbar}{{\text{min}}\left[\Delta E(x) \right]^2 } \propto \frac{\hbar}{\Lambda_0}\frac{1}{\alpha^2(1-\alpha^2)}. \label{eq4-9}
\end{equation}
Then, the time scale $T$ diverges when the parameter $\alpha =\langle \phi_0|v_0 \rangle$ vanishes, that is, the final state $|v_0\rangle$ cannot be obtained when the initial condition is orthogonal to the final sate $|\phi_0\rangle$.
In addition, from Eq. (\ref{eq4-9}), the condition $\alpha=1$ also yields the divergence of the time scale $T$.
However, in such a case, it is not necessary to perform quantum annealing because $|\phi_0\rangle = |v_0\rangle$.
Figure \ref{fig4} shows the functions $a(x)$ and $b(x)$ defined in Eq. (\ref{eq4-5}).
The initial condition $|\psi(0)\rangle =|\phi_0\rangle$ yields the conditions $a(0)=0$ and $b(0)=1$.
As shown in Fig.\ref{fig4}, the state vector $|\psi(t)\rangle$ monotonically changes from the ground state $|\phi_0\rangle$ of the initial Hamiltonian to the first-principal component $|v_0\rangle$ of the information matrix $A$.

\section{\label{sec5} Power series of eigenvectors}
It is interesting to note the power series of eigenvalues of the present way.
We expand the state vector $|\psi(t)\rangle $ as 
\begin{equation}
|\psi(t)\rangle =\sum_{n=0}^{\infty}\left(\frac{t}{T}\right)^n |f_n\rangle, \label{eq5-1}
\end{equation}
using the set of vectors $\{|f_n\rangle \}$ which are time-independent.
Here the vectors $\{|f_n\rangle \}$ are not necessarily orthogonal.
Thus, using Eqs. (\ref{eq3-2}), (\ref{eq3-3}), and (\ref{eq5-1}), we obtain the relations
\begin{equation}
|f_0\rangle =|\phi_0\rangle, \quad |f_1\rangle =\frac{T}{i\hbar}\mathcal{H}_0|f_0\rangle, \label{eq5-2}
\end{equation}
and
\begin{equation}
|f_n \rangle =\frac{T}{i\hbar}\frac{1}{n}\left[ \mathcal{H}_0|f_{n-1}\rangle - (G+\mathcal{H}_0)|f_{n-2}\rangle \right]; \quad (n\geq 2), \label{eq5-3}
\end{equation}
for the initial condition $|\psi(0)\rangle=|\phi_0\rangle$.
Then, at time $t=T$, we obtain the final state $|\psi(T)\rangle$ as
\begin{equation}
|\psi(T)\rangle =\sum_{n=0}^{\infty} |f_n\rangle. \label{eq5-4}
\end{equation}
Unfortunately, we cannot find the general expressions of $|f_n\rangle$.
However, the present expression shows the mixture of states as
\begin{equation}
|\psi(0)\rangle =|\phi_0\rangle \to |\psi(T)\rangle =\sum_{n=0}^{\infty} |f_n\rangle. \label{eq5-5}
\end{equation}
This result may be interesting from the viewpoints of entanglements.
Especially, the recurrence formula shown in Eqs. (\ref{eq5-2}) and (\ref{eq5-3}) is time-independent.
Thus, it is useful for such numerical calculations as image analyses shown in Fig. {\ref{fig2}}.

\section{\label{sec6} Summary and discussions}
In the present study, we demonstrate the possibility to perform the singular value decomposition using the quantum annealing method.
The key idea is to transform the sign of the final Hamiltonian as shown in Eq. (\ref{eq3-1}).
Furthermore, we have investigated the physical background of the present application, namely, the eigenvalue profile and the rigorous expansions.
Finally, we have obtained the applicable time-scale estimated by the energy gap shown in Eq. (\ref{eq4-8}).

The present study suggests an important application of quantum annealing to big data analysis.
Especially, the principal component analysis is important to analyze big data with projecting the data onto lower (at most one- or two-)dimensional data.
The present method will be useful for such statistical physics of information as analyzing big data.

Through this study, we just show a possible way of the singular value decomposition using quantum annealing.
The present discussion is based on the scheme of traditional theory of singular value decomposition (as shown in Sec. \ref{sec2}), and then, our use of quantum annealing does not provide a new kind of matrix decomposition.
However, we expect that the present method will provide new aspects both of singular value decomposition and quantum annealing.
Further studies on the application for quantum physics such as quantum entanglement will propose a new kind of application of quantum annealing.

Finally, we mention the relations between the initial Hamiltonians of typical quantum annealing and that of the present method.
In typical quantum annealing, it focuses on the ground state of classical complex systems $\mathcal{H}_1^{\text{typical}}$ such as spin-glasses.
Then, quantum effects are included in the off-diagonal elements of the initial Hamiltonian $\mathcal{H}_0^{\text{typical}}$.
On the other hand, in the present method, the target Hamiltonian matrix $\mathcal{H}_0=-G$ is, generally, not diagonal, because the matrix $G=A^{\dagger}A$ corresponds to the variance-covariance matrix.
Thus, even if the initial Hamiltonian is defined by such a diagonal matrix as shown in Eq.(\ref{eq3-4}), the state vector $|\psi(t)\rangle$ changes toward to the ground state of $-G$.

\begin{acknowledgments}
One of the authors (Y.H.) is partially supported by the Grants-in-Aid for Young Scientists (B) (Grant No. 26800205) from Japan Society for the Promotion of Science (JSPS).
The authors would like to thank the referees for their useful comments.
\end{acknowledgments}

\bibliographystyle{model1a-num-names}
%\bibliography{<your-bib-database>}

\begin{thebibliography}{00}
\bibitem{1}
T. Kadowaki and H. Nishimori, Phys. Rev. E {\bf 58}, 5355 (1998) .
\bibitem{2}
B. Apolloni, C. Carvalho, and D. de Falco, Stochastic Processes and their Applications {\bf 33}, 233 (1989).
\bibitem{3}
A. B. Finnila, M.A. Gomez, C. Sebenik, C. Stenson, and J.D. Doll, Chem. Phys. Lett. {\bf 219}, 343 (1994).
\bibitem{4}
E. Farhi, J. Goldstone, S. Gutmann, J. Lapan, A. Lundgren and D. Preda, Science {\bf 292}, 472 (2001).
\bibitem{5}
A. Das and B.K. Chakrabarti {\it Quantum Annealing and Related Optimization Methods}, Lecture
Note in Physics (Springer, Verlag, 2005).
\bibitem{6}
G.E. Santoro and E. Tosatti, J. Phys. A {\bf 39}, R393 (2006).
\bibitem{7}
A. Das and B.K. Chakrabarti, Rev. Mod. Phys. {\bf 80} 1061 (2008).
\bibitem{8}
J. Brooke, D. Bitko, T. F. Rosenbaum, and G. Aeppli, Science {\bf 284}, 779 (1999).
\bibitem{9}
P. Ray, B. K. Chakrabarti, and A. Chakrabarti, Phys. Rev. B {\bf 39}, 11828 (1989).
\bibitem{10}
S. Suzuki, J. Inoue and B. K. Chakrabarti,  {\it Quantum Ising Phases and Transitions in Transverse Ising Models}, Lecture Notes in Physics 862 (Springer, Heidelberg, 2013)
\bibitem{11}
H. Chen, X. Kong, B. Chong, G. Qin, X. Zhou, X. Peng, and J. Du, Phys. Rev. A {\bf 83}, 032314 (2011).
\bibitem{12}
M. Steffen, Wim van Dam, T. Hogg, G. Breyta, and I. Chuang, Phys. Rev. Lett. {\bf 90}, 067903 (2003).
\bibitem{13}
X. Peng, Z. Liao, N. Xu, G. Qin, X. Zhou, D Suter, and J. Du, Phys. Rev. Lett. {\bf 101}, 220405 (2008).
\bibitem{14}
J. Du, N. Xu, X. Peng, P. Wang, S. Wu, and D. Lu, Phys. Rev. Lett. {\bf 104}, 030502 (2010).
\bibitem{15}
K. Kurihara, S. Tanaka, and S. Miyashita, Proceedings of the 25th Conference on Uncertainty in Artificial Intelligence, 2009.
\bibitem{16}
I. Sato, S. Tanaka, K. Kurihara, S. Miyashita and H. Nakagawa, Neurocomputing {\bf 121}, 523 (2013).
\bibitem{17}
H. Nishimori, {\it ``Statistical Physics of Spin Glasses and Information Processing: An Introduction''} (Oxford University Press, Oxford, 2001)
\bibitem{18}
E. T. Jaynes, Phys. Rev. {\bf 106}, 620 (1957).
\bibitem{19}
P. Ruj\'{a}n, Phys. Rev. Lett. {\bf 70}, 2968 (1993).
\bibitem{20}
 I. T. Jolliffe, {\it Principal Component Analysis}, Springer Series in Statistics, 2nd ed. (Springer, NewYork, 2002).
\bibitem{21}
C. H. Lee, Y. Yamada, T. Kumamoto, H. Matsueda, J. Phys. Soc. Jpn. {\bf 84}, 013001 (2015).
\bibitem{22}
H. Matsueda, C. H. Lee, Y. Hashizume, arXiv:1403.6259 (2014). 
\bibitem{23}
Y. Imura, T. Okubo, S.Morita, and K. Okunishi, J. Phys. Soc. Jpn. {\bf 83}, 114002 (2014).
\bibitem{24}
H. Matsueda, Phys. Rev. E {\bf 85}, 031101 (2012).  
\bibitem{25}
H. Matsueda, arXiv:1106.5624 (2011). 
\bibitem{26}
M. S. Sarandy, L.-A. Wu and D. A. Lider, Quantum Information Processing {\bf 3}, 331 (2004)
\end{thebibliography}

\end{document}